\documentclass[fleqn,twoside]{article}
\usepackage{espcrc2}
\usepackage{amssymb}
\usepackage{latexsym}

\usepackage{graphicx}
\usepackage[figuresright]{rotating}

\def\bc{\begin{center}}
\def\ec{\end{center}}
\newcommand{\be}{\begin{equation}}
\newcommand{\ee}{\end{equation}}

\newcommand{\MSbar}{\overline{\mathrm{MS}}}

\newcommand{\nn}{\nonumber}

\newcommand{\acpi}{\mfrac{\alpha_{s}}{4\pi}}

\newcommand{\beq}{\begin{equation}}
\newcommand{\eeq}{\end{equation}}
\newcommand{\beqs}{\begin{equation*}}
\newcommand{\eeqs}{\end{equation*}}
\newcommand{\beqn}{\begin{eqnarray}}
\newcommand{\eeqn}{\end{eqnarray}}
\newcommand{\bea}{\begin{eqnarray}}
\newcommand{\eea}{\end{eqnarray}}
\newcommand{\beqns}{\begin{eqnarray*}}
\newcommand{\eeqns}{\end{eqnarray*}}
\newcommand{\mfrac}[2]{\frac{\textstyle #1}{\textstyle #2}}

\def\vdir{v\kern-5.75pt\raise0.15ex\hbox{${\scriptstyle /}$}}
\def\pdir{p\kern-7.8pt\raise0.2ex\hbox{\Big{/}}}
\def\ddir{D\kern-7pt\raise0.2ex\hbox{\big{/}}}
\def\partdir{\partial\kern-7.6pt\raise0.25ex\hbox{/}}
\def\ddirp{D_{\kern-2.75pt\perp}\kern-11pt\raise0.2ex\hbox{\big{/}}\kern+4.5pt}

\newcommand{\AmS}{{\protect\the\textfont2
  A\kern-.1667em\lower.5ex\hbox{M}\kern-.125emS}}

\title{HQET with chiral symmetry on the lattice
\thanks{This work has been  supported in part by the EU IHP under HPRN-CT-2000-00145
Hadrons/LatticeQCD.}}

\author{
Damir Be\'cirevi\'c\address{Laboratoire de Physique Th\'eorique (B\^at.210),
Universit\'e Paris Sud, Centre d'Orsay, 91405 Orsay-Cedex, France and CERN, Theory Division, CH-1211 Geneva 23, Switzerland.} and 
Juan Reyes\address{Dip. di Fisica, Univ. di Roma ``La Sapienza'' and
INFN-Sezione di Roma I, Piazzale A. Moro 2, I-00185 Roma, Italy.}}
\begin{document}

\begin{abstract}
\mbox{}\\[-0.5cm]
We show that by  combining the static heavy quark action with the Neuberger action for the light quark, the renormalisation 
 of the heavy-light bilinear and four-quark operators, computed on the lattice, becomes highly 
simplified: all the heavy-light bilinears get renormalised by a single multiplicative
constant, whereas the renormalisation of the complete set of parity even $\Delta B=2$ four-quark 
operators involves only four independent constants. The relevant (matching) constants are computed
at NLO in perturbation theory and are presented here.  
\end{abstract}
% typeset front matter (including abstract)
\maketitle
Computation of the $B$-meson decay constants and the corresponding 
``bag parameters" by using lattice QCD is very demanding and beyond reach to 
the currently available computing facilities. In such a situation, the results 
for these quantities obtained 
in the static limit, $m_Q\to \infty$, are 
particularly helpful. They can be combined with those  
obtained by using the propagating heavy quarks with $m_Q < m_b$, to interpolate 
in the inverse heavy quark mass and thus extracting the physically interesting 
quantities at $m_Q=m_b$ (see eg. ref.~\cite{p1}). However, after combining the static
HQET action with the standard Wilson light quark Lagrangian on the lattice, 
the explicit breaking of the chiral symmetry generates many problems in 
the renormalisation procedure of the local heavy-light operators. In particular, the mixing 
pattern in the renormalisation of the four-quark operators becomes very complicated.  
The extra mixing is a lattice artifact and turns out to be uncomfortably large for 
the bare lattice couplings actually used in practice. For that reason, disentangling 
an operator that we want to match to its continuum counterpart is difficult and is prone 
to the additional systematic and statistical errors. Furthermore,  when working with 
Wilson light quarks, the frequent 
appearance of exceptional configurations prohibits studying the quarks lighter than 
a half of the strange quark mass. The above problems can be overcome if the chiral symmetry 
of the light quark is exactly preserved on the lattice. In recent years it became 
evident that the {\it overlap fermion action}~\cite{p2} indeed preserves the chiral 
symmetry on the lattice without giving up any other symmetry.

In what follows, we will show that the combination of the Neuberger (overlap) light and the 
static heavy quark action indeed simplifies the renormalisation procedure 
of the composite operators. We will then present the expressions for the renormalisation 
constants for the heavy-light bilinears, as well as for the $\Delta B=2$ four-quark operators, 
that we derived at one-loop (NLO) in perturbation theory.

\section{Action and symmetries}
The action that we are interested in, has the following form:
\beqn \label{action}
&&{\cal S} = {\cal S}_{\rm YM}+ {\cal S}_{\rm light} + {\cal S}_{\rm heavy}\,.
\eeqn 
For the discretised version of the Young-Mills part of the action ${\cal S}_{\rm YM}$ we take the 
standard Wilson plaquette action, whereas for the ${\cal S}_{\rm heavy}$ part we adopt the static HQET action with  
the backward derivative prescription of ref.~\cite{p3}, i.e. 
\beqns
S_{\rm heavy}\!\!\!\!\!\!&=\!\!\!&  \sum_{n} \left\{ \bar h^{(+)}(n) \left[ h^{(+)}(n) - U_0(n-\hat 0)^\dagger
h^{(+)}(n-\hat 0)\right]\right.\\
   &&\left. - \bar h^{(-)}(n)\left[  U_0(n)h^{(-)}(n+\hat 0) -
   h^{(-)}(n)\right]\right\},
\eeqns
where $h(n)$ is the static heavy quark field, and $U_0(n)$ is the link variable in the temporal direction. 
For the light quark we take the Dirac Lagrangian \`a la Neuberger~\cite{p2},
\[
D_N={1\over a}\rho\left[1+{X\over \sqrt{X^\dagger X}}\right]\:,
\quad X=D_W-{\rho \over a}\ ,
\]
where $D_W$ is the Wilson Dirac operator, $2 D_W=\gamma_\mu (\nabla_\mu^\ast+\nabla_\mu)- a\nabla^\ast_\mu\nabla_\mu$.

On the finite lattice, the action~(\ref{action}) is invariant with respect to:\\
{\bf 1.} \underline{Chiral symmetry} transformations ($\chi$S)~\cite{p4}
\beqn
\psi(x)&\to&  i\gamma_5 \psi(x)\,,\quad 
\bar \psi(x) \to \bar\psi(x) i\,(1-{D_N\over \rho}) \gamma_5;
\label{sym_chi}
\nn
\eeqn
{\bf 2.} \underline{Heavy quark spin symmetry} transformations (HQS), 
of which we will need the following ones
\beqn\label{sym_spin}
&&h^{(\pm)}(x)\to \frac{1}{2}\epsilon_{ijk}\gamma_j\gamma_k h^{(\pm)}(x)\;,\quad
\nn\\
&&\bar h^{(\pm)}(x)\to  -\bar h^{(\pm)}(x)\frac{1}{2}\epsilon_{ijk}\gamma_j\gamma_k\;,\quad
(i=1,2,3);\nn
\eeqn
{\bf 3.} \underline{Discrete O(3) symmetry}, of which we will need just a 
rotation by $\pi/2$ about the $i^{th}$ axis ($x_i\to x_i$, $x_{j\neq i} \to \epsilon_{ijk} x_k$):
\beqn\label{sym_SO3}
&&\hspace{-0.5cm}\psi(x)\; h^{(\pm)}(x)\to {1-\frac{1}{2}\epsilon_{ijk}\gamma_j\gamma_k\over
\sqrt{2}}\psi(x)\;h^{(\pm)}(x),\nn\\
&&\hspace{-0.5cm}\bar\psi(x)\;\bar h^{(\pm)}(x)\to \bar\psi(x)\;\bar
h^{(\pm)}(x){1+\frac{1}{2}\epsilon_{ijk}\gamma_j\gamma_k\over
\sqrt{2}}.\nn
\eeqn

With the above symmetry properties in hands, we now show that all the heavy-light bilinears
\beqn
O_{\Gamma}=\bar h\, \Gamma\psi,\quad \Gamma &\in&
\{\mathbf{1},\gamma_5,\gamma^\mu,\gamma^\mu\gamma_5,{i\over 
2}[\gamma^\mu,\gamma^\nu]\}\cr
O_\Gamma &\in&
\{S,P,V^\mu,\; A^\mu,\qquad T^{\mu\nu}\}, \nn
\eeqn
renormalise by a single renormalisation constant. 
A consequence of the heavy quark field equation, 
$\bar h\gamma_0=\bar h$, is that $V^0=S$, $A^0=P$, and the tensor density is not independent bilinear 
($T^{0j}=V^j$,  $T^{ij}=\varepsilon^{ijk}A^k$).  
Under the $\chi$S transformation, we see that 
$S \to iP$, $P\to iS$, and $V^j\to iA^j$, $A^j\to iV^j$, so that the multiplicative renormalisation constants satisfy 
$Z_A=Z_V$, $Z_S=Z_P$. On the other hand, under the HQS transformations, we have 
$S \to iA^j$, $P\to iV^j$, and $V^j\to iP$, $A^j\to iS$, implying that 
 $Z_S=Z_A$ and $Z_V=Z_P$. We thus arrive at the wanted result that
\beqn\label{2f}
&&Z_A= Z_V=Z_S=Z_P\equiv Z(a\mu)\ .
\eeqn
With the notation, $O_{\Gamma_1 \Gamma_2}=(\bar h^{(+)\,\alpha} \Gamma_1q^{\alpha})
              (\bar h^{(-)\,\beta}\Gamma_2q^{ \beta})$, we chose the following basis of
 parity conserving $\Delta B=2$ operators (in HQET):
\beqn
O_{\Gamma \Gamma}&\!\!\!\!\in\!\!\!& \{O_{VV+AA}, O_{SS+PP},  O_{VV-AA}, 
O_{SS-PP}\}.\nn
\eeqn
 Without symmetries discussed above, all entries of the $4\times 4$ renormalisation matrix are  
non-zero and independent from one another. HQS, together with $O(3)$, provides relations among the entries, 
resulting in the following structure~\cite{p5}):
\be\label{Z_PC}
Z=
\left(\matrix{ {Z_{11}} & 0 & {{Z}_{13}} & 2\,
   {{Z}_{13}} \cr \frac{-{Z_{11}} + {Z_{22}}}
   {4} & {Z_{22}} & {{Z }_{23}} & -{{Z
        }_{13}} - 2\,{{Z }_{23}} \cr {{Z}_
    {31}} & {{Z }_{32}} & {Z_{33}} & {Z_{34}} \cr 
    \frac{2\,{{Z }_{31}} - {{Z}_{32}}}{4} & 
    \frac{-{{Z }_{32}}}{2} & \frac{{Z_{34}}}{4} & 
    {Z_{33}}\nn \cr \nn }\nn \right)\nn , \nonumber
\ee
as explicitly verified in perturbation theory with Wilson light quarks~\cite{p6}. 
$8$ independent entries get reduced to only $4$, after applying the $\chi$S transformations. Indeed, we verify that
\beqn
O_{VV+AA} \leftrightarrow -\,  O_{VV+AA} ,&&\!\!
O_{SS+PP} \leftrightarrow -\,  O_{SS+PP},\nn\\ 
O_{VV-AA} \leftrightarrow +\,  O_{VV-AA} ,&&\!\!
O_{SS-PP} \leftrightarrow +\,  O_{SS-PP}, \nonumber
\eeqn
which finally brings us to the form,
\be\label{Z4f_1}
%\left(\matrix{ O_{VV+AA} \cr O_{SS+PP} \cr O_{VV-AA} \cr
%O_{SS-PP}}\right)^{R}
%=
Z=\left(\matrix{ {Z_{11}} & 0 & 0 & 0 \cr \frac{-{Z_{11}} + {Z_{22}}}
   {4} & {Z_{22}} & 0 &  \cr 0 & 0& {Z_{33}} & {Z_{34}} \cr 
    0 & 0 & \frac{{Z_{34}}}{4} & 
    {Z_{33}} \cr  }\right).
%    \cdot\left(
%    \matrix{ O_{VV+AA}\cr O_{SS+PP}\cr O_{VV-AA}\cr O_{SS-PP}}\right)^{B}
\ee

\section{Perturbative matching}
In this section we present results of the procedure in which we match the lattice regularised operators 
with their continuum counterparts, renormalised in the ${\MSbar}$(NDR) renormalisation scheme. 
Details of the calculation will be presented in ref.~\cite{p5}. Here we only spill out the results.

For the renormalisation constant relevant to the bilinear operators [cf. eq.(\ref{2f})], we obtain
\begin{eqnarray} 
Z^{\MSbar}(\mu\,a) &=& 1+\acpi {4\over 3}\ \left[
{5\over 4}-d_\Sigma-e-d_H\right.\nn\\
&&+ \left.{3\over2}\,\ln(\mu^2 a^2)\right] \:.\nn
\end{eqnarray}
where the constant $e=24.48059730$, comes from the self energy of the static quark leg~\cite{p3}, 
$d_\Sigma(\rho)$ comes from the light quark self energy~\cite{p7}, and $d_H$ is the vertex 
contribution that has not been calculated before. The values for $d_{\Sigma,H}(\rho)$ are listed in tab.~\ref{tabela}, for 
three specific values of parameter $\rho$.

For the renormalisation constants of the four fermion operators our results read:
\beqns
Z_{11}&=&1+\acpi \left[\frac{7}{3} - \frac{c}{3} - \frac{10\, {d_H}}{3} + \
\frac{{d_S}}{3} - \frac{4\, {d_\Sigma }}{3}\right.\\
 &-&\left.  \frac{4\, 
      e}{3} + \frac{2\, {d_\xi}}{3} + 4\, \log (a^2 {\mu }^2)\right]\\
Z_{21}&=&\acpi\left[- \frac{5}{36}   + \frac{c}{4} + \
\frac{{d_H}}{2} - \frac{{d_S}}{36} - \frac{2\, \
{d_V}}{9}\right.\\
 &-& \left. \frac{{d_\xi}}{6} - \frac{2\, \log (a^2 {\mu \
}^2)}{3}\right]\\
Z_{22}&=&1+\acpi \left[\frac{16}{9} + \frac{2\, 
      c}{3} - \frac{4\, {d_H}}{3} + \frac{2\, {d_S}}{9}\right.\\
  &-&\left. \
\frac{8\, {d_V}}{9} - \frac{4\, {d_\Sigma }}{3} - \frac{4\, 
      e}{3} + \frac{4\, \log (a^2 {\mu }^2)}{3}\right]\\
Z_{33}&=&1+\acpi\left[\frac{41}{12} + \frac{c}{6} - \frac{7\, {d_H}}{3} - \
\frac{{d_V}}{6} - \frac{4\, {d_\Sigma }}{3}\right.\\
 &-&\left. \frac{4\, 
      e}{3} + \frac{7\, {d_\xi}}{6} + \frac{7\, \log (a^2 {\mu \
}^2)}{2}\right]\\
Z_{34}&=&\acpi\left[\frac{1}{2} + c + 2\, {d_H} - {d_V} - {d_\xi}\right.\\
 &-&\left. 
  3\, \log (a^2 {\mu }^2)\right]\\
Z_{43}&=&\acpi\left[\frac{1}{8} + \frac{c}{4} + \frac{{d_H}}{2} - \
\frac{{d_V}}{4} - \frac{{d_\xi}}{4}\right.\\
 &-&\left. \frac{3\, \log (a^2 \
{\mu }^2)}{4}\right]\\
Z_{44}&=&1+\acpi\left[\frac{41}{12} + \frac{c}{6} - \frac{7\, {d_H}}{3} - \
\frac{{d_V}}{6} - \frac{4\, {d_\Sigma }}{3}\right.\\
 &-&\left. \frac{4\, 
      e}{3} + \frac{7\, {d_\xi}}{6} + \frac{7\, \log (a^2 {\mu \
}^2)}{2}\right]
\eeqns
where the constant $c=4.52575660$ (less precise value was first obtained in ref.~\cite{p8}), 
while the constant $d_\xi= -4.79200957$. 
The values of $d_{S,V}(\rho)$ were already computed in ref.~\cite{p7}, which we checked and 
confirm here. The values are listed in tab.~\ref{tabela}. 
Notice that our results indeed verify the symmetry relation, 
\be
Z_{21}={Z_{22}-Z_{11}\over 4},\quad Z_{33}=Z_{44},\quad Z_{43}={Z_{34}\over 4}.\nn
\ee
\begin{table}[t!]
\bc
\begin{tabular}{cccc}
\hline \hline
{$\rho$}&$1.0$ &$1.2$&$1.4$  \\
\hline \hline
$d_\Sigma(\rho)$  & -31.33861723& -23.20304037 & -17.47396963\\
$d_H(\rho)$       & 0.55183709  & 0.59728235   & 0.64838696\\
$d_S(\rho)$  	  & 1.46989129& 2.02526759   & 2.55134784\\
$d_V(\rho)$  	  & 0.03924115& 0.04665160   & 0.05600630\\
\hline
\end{tabular}
\ec
\caption{\small  Numerical values of  $d_\Sigma(\rho)$, $d_H(\rho)$, $d_S(\rho)$ and
$d_V(\rho)$ for
three values of $\rho$ (see \cite{p5} for other values of $\rho$).}  
\label{tabela}
\vspace*{-5mm}
\end{table}

\section{Concluding remarks}
Our proposal to combine the static HQET with the overlap light quark is very rewarding 
for the renormalisation procedure. It particularly simplifies the renormalisation of
the phenomenologicaly important $\Delta B=2$ operators. In practice, the use of HQET on the 
lattice suffers from the poor signal-to-noise ratio. This problem was recently circumvented 
by replacing $U_0(n)\to U_0^{fat}(n)$ (fat-link) in ${\cal S}_{heavy}$~\cite{p9}.
Empirically, that replacement results in the statistical accuracy in correlation functions, 
comparable to what one has in lattice QCD. The implementation of that 
replacement in the above results will be presented elsewhere.


\begin{thebibliography}{9}

\bibitem{p1}
D.~Becirevic {\it et al.}, JHEP {\bf 0204} (2002) 025.
% [hep-lat/0110091].
%%CITATION = HEP-LAT 0110091;%%


\bibitem{p2}
H.~Neuberger,
Phys.\ Lett.\ B {\bf 417} (1998) 141.
% [hep-lat/9707022].
%%CITATION = HEP-LAT 9707022;%%


\bibitem{p3}
E.~Eichten and B.~Hill,
Phys.\ Lett.\ B {\bf 240} (1990) 193.
%%CITATION = PHLTA,B240,193;%%


\bibitem{p4}
M.~Luscher,
Phys.\ Lett.\ B {\bf 428} (1998) 342.
% [hep-lat/9802011].
%%CITATION = HEP-LAT 9802011;%%


\bibitem{p5}
D.~Be\'cirevi\'c and J.~Reyes, in preparation.

\bibitem{p6}
M.~Di Pierro and C.~T.~Sachrajda,
Nucl.\ Phys.\ B {\bf 534} (1998) 373;
% [hep-lat/9805028]
%%CITATION = HEP-LAT 9805028;%%
V.~Gimenez and J.~Reyes,
{\it ibid} {\bf 545} (1999) 576. 
%[hep-lat/9806023].
%%CITATION = HEP-LAT 9806023;%%

\bibitem{p7}
C.~Alexandrou {\it et al.}, 
Nucl.\ Phys.\ B {\bf 580} (2000) 394,
%[arXiv:hep-lat/0002010].
%%CITATION = HEP-LAT 0002010;%%
S.~Capitani and L.~Giusti,
Phys.\ Rev.\ D {\bf 62} (2000) 114506.
%[arXiv:hep-lat/0007011].
%%CITATION = HEP-LAT 0007011;%%

\bibitem{p8}
J.~M.~Flynn, O.~F.~Hernandez and B.~R.~Hill,
Phys.\ Rev.\ D {\bf 43} (1991) 3709.
%%CITATION = PHRVA,D43,3709;%%

%\cite{Gonzalez-Arroyo:1981ce}
%\bibitem{constants}
%A.~Gonzalez-Arroyo and C.~P.~Korthals-Altes,
%``Asymptotic Freedom Scales For Any Lattice Action,''
%Nucl.\ Phys.\ B {\bf 205} (1982) 46.
%%CITATION = NUPHA,B205,46;%%
%G.~Burgio, S.~Caracciolo and A.~Pelissetto,
%``Algebraic algorithm for the computation of one-loop Feynman diagrams in  lattice QCD with Wilson fermions,''
%Nucl.\ Phys.\ B {\bf 478} (1996) 687.
%[arXiv:hep-lat/9607010].
%%CITATION = HEP-LAT 9607010;%%

\bibitem{p9}
M.~Della Morte {\it et al.}, hep-lat/0307021.
%%CITATION = HEP-LAT 0307021;%%


\end{thebibliography}
\end{document}